\documentclass{article}

\usepackage{arxiv}

\usepackage[utf8]{inputenc} 
\usepackage[T1]{fontenc}    
\usepackage[superscript]{cite}
\usepackage{hyperref}       
\usepackage{url}            
\usepackage{booktabs}       
\usepackage{amsfonts}       
\usepackage{nicefrac}       
\usepackage{microtype}      
\usepackage{lipsum}
\usepackage{graphicx}
\usepackage{authblk}       
\usepackage{textgreek}
\usepackage{multirow}

\title{Comparing transient oligonucleotide hybridization kinetics using DNA-PAINT and optoplasmonic single-molecule sensing on gold nanorods}

\author[1,2]{Narima Eerqing}
\author[1,2]{Sivaraman Subramanian}
\author[2]{Jes\'us Rubio}
\author[1,2]{Tobias Lutz}
\author[1,2]{Hsin-Yu Wu}
\author[2,3]{Janet Anders}
\author[1,2]{Christian Soeller}
\author[1,2]{Frank Vollmer}

\affil[1]{Living Systems Institute, University of Exeter,
  Stocker Road, Exeter, EX4 4QD, United Kingdom }
\affil[2]{Department of Physics and Astronomy, University of Exeter, Stocker Road, Exeter EX4 4QL, United Kingdom}
\affil[3]{Institut f\"ur Physik und Astronomie, University of Potsdam, 14476 Potsdam, Germany}

\begin{document}
\maketitle
\begin{abstract}
We report a comparison of two photonic techniques for single-molecule sensing: fluorescence nanoscopy and optoplasmonic sensing. As the test system, oligonucleotides with and without fluorescent labels are transiently hybridized to complementary ‘docking’ strands attached to gold nanorods. Comparing the measured single-molecule kinetics helps to examine the influence of fluorescent labels as well as factors arising from different sensing geometries. Our results demonstrate that DNA dissociation is not significantly altered by the fluorescent label, while DNA association is affected by geometric factors in the two techniques. These findings open the door to exploiting plasmonic sensing and fluorescence nanoscopy in a complementary fashion, which will aid in building more powerful sensors and uncovering the intricate effects that influence the behavior of single molecules. 
\end{abstract}

Biomolecules inhabit a microbiological environment sufficiently complex for their state to vary such that biological processes operate far from thermodynamic equilibrium. As a result, information extracted with ensemble measurements can be insufficient to predict the behavior of biomolecules, since ensemble averages may be far from the state of any single molecule in the system.\cite{Lenn2012, Miller2017} The alternative is to probe biomolecular activity directly by using single-molecule techniques,\cite{Capitanio2013,Yu2011,Kufer2009,Hinterdorfer2006} which do not only provide mean values of physical observables but also more detailed statistics in the form of probability distributions. A plethora of photonics-based techniques for single-molecule detection has emerged in the past decade. For example, single-molecule localization microscopy (SMLM)\cite{Rust2006,Betzig2006,Hess2006} has overcome the diffraction limit, enabling the reconstruction of images with great precision via temporal modulation and the accumulation of single-molecule detection events. Examples include photo-activated localization microscopy (PALM),\cite{Betzig2006} stochastic optical reconstruction microscopy (STORM),\cite{Rust2006} and DNA-based point accumulation for imaging in nanoscale topography (DNA-PAINT).\cite{Jungmann2014,Schnitzbauer2017,Lin2020} On the other hand, single-molecule detection that makes use of noble metal nanostructures has attracted great attention owing to the exceptional sensitivity arising from localized surface plasmon resonance (LSPR).\cite{Sepulveda2009,Zhang2018w} The local enhancements of optical near-fields around gold nanorods (GNRs) have enabled particularly high detection sensitivities for biomolecules without the need for fluorescent labels.\cite{Malinsky2001,Haes2004,Nusz2009} Among the family of LSPR-based techniques, the combination of whispering-gallery-mode (WGM) sensing with plasmonic nanorods (optoplasmonic sensing) has shown a very high detection sensitivity even down to small biomolecules such as short DNA oligonucleotides.\cite{Baaske2014,Baaske2016,Vincent2020}

Despite this steady progress, single-molecule techniques suffer from different drawbacks that limit the information one can extract. For example, it is well-known that fluorescent labels may affect the kinetics and dynamics of biomolecular systems,\cite{Dietz2019,Peterson2019,Zimmers2019} while LSPR-based approaches use local field enhancements that are not uniform across a plasmonic nanostructure, and the local surface geometry and heterogeneity of plasmonic nanoparticles can play a significant role in the observed statistics.\cite{Willets2019,Byers2014,Kaushal2020} In view of this, it is desirable to combine the information arising from different methods when applied to the same molecular system. But, to guarantee the validity of such an approach, current single-molecule techniques must be compared and cross-validated.

\begin{figure}[t]
    \centering
    \includegraphics[width=0.95\textwidth]{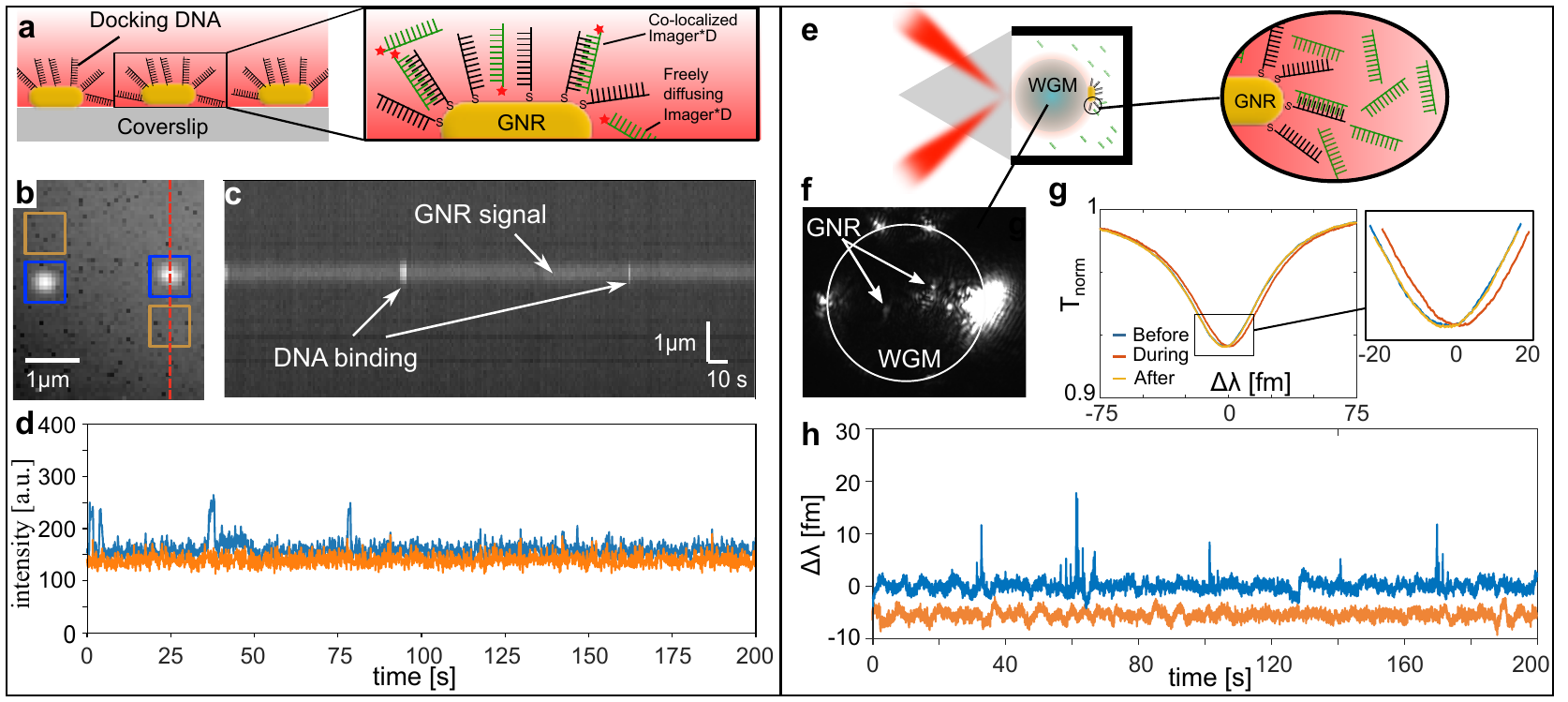}
    \caption{Comparison between DNA-PAINT (a-d) and WGM (e-h) methods. (a) Schematic of DNA-PAINT experiment on gold nanorods (GNRs). GNRs with LSPR $\sim 750$\,nm (yellow) are coated with DNA docking strands P1 or T22 (black) and immobilized on the surface of a coverslip (grey). The zoomed view shows a single immobilized GNR. Imager strands (green) labeled with flourophore (ATTO655/DY782, red star) are added and excited. (b) Microscope image showing the signals co-localized with the GNRs (blue box) and adjacent to the GNRs (orange box) positions. (c) A space-time plot of the red line through one GNR in (b). The GNRs provide a constant intensity over time and the hybridization of imagers are observed as short bursts of increased intensity. (d) Typical fluorescence time traces, with orange and blue lines showing the intensity signal at the locations indicated in (b). (e) Schematic of WGM sensing approach. A laser beam is focused on the surface of the prism and coupled into a glass microsphere. GNRs are then immobilized on the surface of the microsphere. The zoom-in shows docking strands P1 or T22 (black) immobilized on the GNRs surface and imager strands diffusing freely in solution. (f) Microscope image of the glass microsphere with GNRs immobilized on its surface. The scattering from the GNRs is observed as bright spots. (g) An example of wavelength shift caused by the hybridization of the imager and docking strands on the sensor. The blue, orange, and yellow curves represent the WGM mode position before, during, and after a single molecule transient event. (h) Typical detrended time trace of WGM resonance wavelength showing background signal (orange) and signals obtained for DNA hybridization (blue). All experiments were conducted at room temperature ($22.5\,\rm{^o C}$). }
    \label{fig:F1}
\end{figure}

Here we compare the detection of DNA hybridization using two established photonic single-molecule techniques: DNA-PAINT and WGM-based optoplasmonic sensing (see schematic in Figure\,\ref{fig:F1}). We employ two sets of DNA strands, a 13-mer termed P1, a 22-mer termed T22 and their respective complementary strands (see Table\,\ref{tab:T1}). The complementary strands (termed imagers) are labeled (ImP1*D with ATTO 655 and ImT22*D with DY782) for use in DNA-PAINT, and unlabeled (ImP1* and ImT22*) for use with the optoplasmonic sensor. P1 and its imager strand ImP1* are short sequences which have been widely used in DNA-PAINT experiments. \cite{Jungmann2014,Lin2020,Jungmann2010}. T22 is a 22-mer sequence which has been reported to work well in an optoplasmonic sensing platform.\cite{Baaske2014}. In both techniques, the P1 and T22 strands are attached to gold nanorods via thiol linkages and serve as the docking strands, while the imager strands are freely diffusing in solution. The oligonucleotide sequence is chosen to allow for transient hybridization events at room temperature. Within this context, a natural hypothesis is that the kinetics associated with the dissociation of docking and imager strands is independent of the chosen platform. By carrying out the two experiments, DNA-PAINT and WGM, each for two types of DNA sets, such hypothesis can be tested.

\begin{figure}[t]
    \centering
    \includegraphics[width=1\textwidth]{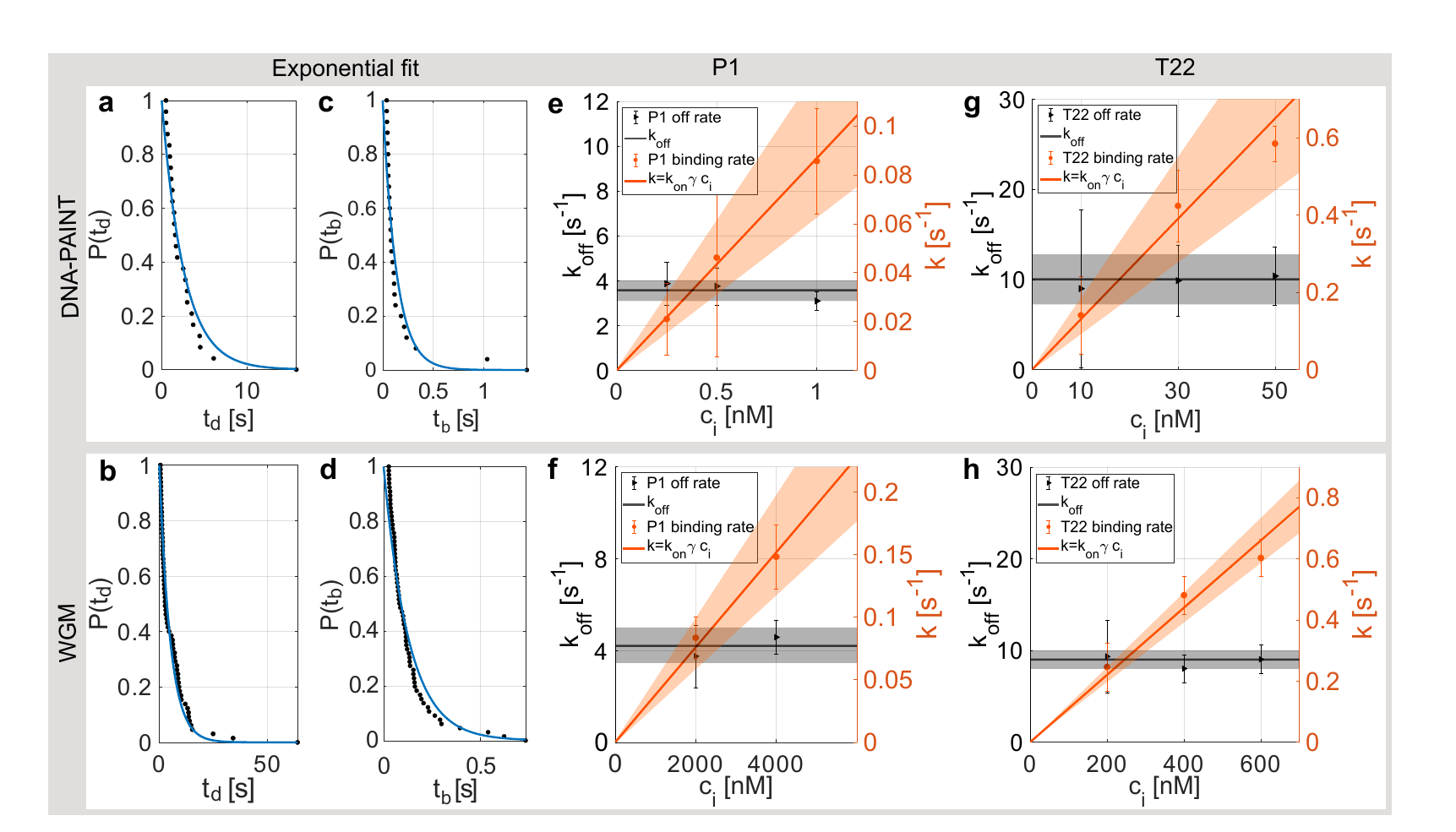}
    \caption{(a,b) Shown are the probabilities P that a binding event has not taken place within an interval $t_d$ for (a) DNA-PAINT with 50nM of ImT22*D and (b) WGM experiments with 200nM of ImT22*, measured (black dots) and fit (blue line). (c,d) Similarly, the probability that a dissociation event has not taken place within an interval $t_b$ is shown in (c) DNA-PAINT with 50nM of ImT22*D and (d) WGM experiment with 200nM of ImT22*. For P1 strands, analogous exponential profiles emerge. (e-h) The single-molecule dissociation rate $k_{\mathrm{off}}$ (black) and the binding rate $k = k_{\mathrm{on}} \gamma c_i$  (orange), both with units of $\rm{s^{-1}}$, are shown for P1 and T22 strands with (e,g) DNA-PAINT and (f,h) WGM sensing, respectively, where $k_{\mathrm{on}}$ is the association rate constant with units of $\rm{M^{-1} s^{-1}}$, $\gamma$ is a numerical factor accounting for the geometric arrangement and accessibility of the docking strands contributing to the signals, and $c_i$ is the molar imager concentration. Our results show that the dissociation rate for the DNA-PAINT and the WGM techniques are the same within error. For binding kinetics, only the product $k_{\mathrm{on}} \gamma$ is retrieved. The uncertainties associated with $k$ and $k_{\rm{off}}$ are given, for each exponential fit, by SE$\,\times\,t_{n-1}$, where SE is the standard error and $t_{n-1}$ is the t-value for $n-1$ degrees of freedom at a $95\%$ level. The uncertainties for the linear fits have been propagated from the former uncertainties.}
    \label{fig:F3}
\end{figure}

Our measurements reveal that the dissociation rates of hybridized DNA strands are indeed approximately the same for both techniques, demonstrating the comparability of the single-molecule information extracted from each platform. In addition, this suggests that the modification of a short strand oligonucleotide with an fluorescent dye at its 5’ end does not significantly affect the dissociation kinetics. On the other hand, the association rates cannot be directly estimated from our measurements. This is due to the surface heterogeneities of the gold nanorods,\cite{Baaske2016} as well as to the variability in the number of strands that are contributing to the signals in each technique. In the following we will detail how the experimental schemes and the data analysis have been carried out, before showing the main results in Fig.~\ref{fig:F3}.

\begin{table}[ht]
\renewcommand{\arraystretch}{1.5}
\centering
\caption{Sequences of ssDNA used for the experiments.}
\label{tab:T1}
\begin{tabular}{lll}
\hline
& \textbf{ssDNA}   & \textbf{Sequence (5'-3') }\\
\hline
\multirow{3}{4em}{Set I} & P1               & [ThiolC6] TTT TAT ACA TCT A                 \\
& ImP1*D      & [Atto655] CTA GAT GTA T                 \\
&ImP1*       & CTA GAT GTA T          \\     
\hline
\multirow{3}{4em}{Set II}& T22         & [ThiolC6] TTT TGA GAT AAA CGA GAA GGA TTG AT                 \\
& ImT22*D     & [DY782] ATC AGT CCT TTT CCT TTA TCT C       (3 mismatched)         \\
& ImT22*      & ATC AGT CCT TTT CCT TTA TCT C     (3 mismatched)     \\     
\hline
\end{tabular}
\end{table}

Figure\,\ref{fig:F1} shows the experimental schemes for both techniques employed in this work. In DNA-PAINT, the thiolated ssDNA are immobilized on to GNRs adsorbed to a glass coverslip as shown in Figure\,\ref{fig:F1}a. The hybridization of the labeled complementary DNA is observed by localization of a fluorescent label attached to the imager DNA using a camera. In the WGM-based optoplasmonic technique, the shift of the resonance wavelength $\lambda$ of WGMs excited in a spherical glass resonator is utilized to monitor the hybridization of the ssDNA. In practice, this involves tracking the changes in the WGM resonance position using a centroid method,\cite{Baaske2016,Dahlin2006} from where we can obtain temporal traces for the wavelength shift $\Delta \lambda$. Similar to DNA-PAINT, a thiolated ssDNA is immobilised on to GNRs adsorbed to the glass resonator surface; see Figure\,\ref{fig:F1}e. In this case, the imager DNA strand does not contain a fluorescent label. The signals are instead obtained due to the plasmonic enhancement provided by the GNRs. \cite{Baaske2014}   

For DNA-PAINT (Figure\,\ref{fig:F1}a), the experiments are carried out on a Nikon Ti-E inverted microscope with a 1.49 NA total internal reflection fluorescence (TIRF) objective (Nikon Apo TIRF 60X oil or Nikon Apo TIRF 100x oil) illuminated using a 647nm (OMicron LuxX 647- 140) or 780 nm (Toptica DL Pro, 100mW) and custom-built optics. The fluorophores are excited via TIR at a coverslip (170 \textmugreek m nominal thickness) and the images are captured real-time by a camera (Andor Zyla 4.2, 10 frames/s or Andor iXON 888, 25 frames/s). The glass coverslips is glued to a chamber cut from acrylic and placed on the microscope stage. The acrylic chamber is used to pipette the various samples required for the experiments. For the WGM technique (Figure\,\ref{fig:F1}e), the experiments are carried out in a custom built prism based setup. Whispering gallery modes in spherical glass resonators (diameter $\sim$ 80 \textmugreek m  fabricated by melting an single mode optical fiber (SMF 28e, Corning) are excited via frustrated total internal reflection at a prism (NSF11, Schott) surface. A microscope setup (10x Olympus) is used to help align the resonator with respect to the prism for efficient coupling. A tunable 780 nm laser source (Toptical TA Pro 780) is scanned at a rate of 50 Hz to capture the WGM spectrum. The position of the resonance peak and the full-width-at-half-maximum (FWHM) are extracted using a custom Labview program. A chamber made of Polydimethylsiloxane (PDMS) is placed around the resonator and is used for injecting samples into the sensor. All experiments were conducted at 22.5 $\rm{^o}$C. 

We carry out both the DNA-PAINT and optoplasmonic sensing experiments for two sets of DNA strands (see Table\,\ref{tab:T1}).  The procedures of the experiments are similar, and contain three steps. First, we immobilize gold nanorods (A12-10-CTAB-750; Nanopartz Inc.) irreversibly on a clean glass surface (a coverslip in case of DNA-PAINT or the spherical resonator surface in case of optoplasmonic sensor) in an acidic aqueous suspension (pH $\approx $1.6, 0.1 nM). This step is run over for 15 min for the GNRs to freely deposit. The chamber is washed thrice to remove unbound GNRs. On average, the diameter of GNRs is 10 nm, and their length is 35 nm with a longitudinal plasmon resonance at 750 nm. In DNA-PAINT, the clean coverslip is pre-functionalized with PLL-g-PEG (Su-Sos) whose role here is to prevent unspecific binding between fluorophore and the coverslip. In the DNA-PAINT setup, the GNRs are excited via TIR at the coverslip and their photoluminescence is visualized real-time by the sCMOS/EMCCD camera. In the optoplasmonic sensing platform, the immobilization of the GNRs are monitored via the shift in WGM resonance wavelength and FWHM. Some of the attached GNRs can also be monitored via the microscope as shown in Figure\,\ref{fig:F1}f.

Next the docking strands (thiolated ssDNA) are immobilised on the GNRs via a mercaptohexyl linker at their 5' end. The procedures are the same for both techniques. The P1 docking strand (see Table\,\ref{tab:T1} for sequences) are immobilised in citrate buffer at pH $\approx$ 3 with 1 M of NaCl. The T22 docking strands are immobilised at the same pH and salt concentration, but with different buffer: 0.02\% wt/wt sodiumdodecylsulfate (SDS) solution. Before the addition of the docking strands into the sample chamber, they are pre-mixed with a solution of a reducing agent (10 \textmugreek l of $10\,\rm{mM}$ tris(2-carboxyethyl)phosphine, TCEP) to cleave the disulfide bond and therefore enable efficient binding of the thiols to the GNRs. The docking strands are then injected into the sample chamber to a final concentration of 1 \textmugreek M and left for 30 mins. Since the docking strands do not contain any fluorescent labels, the binding of the docking strands to the GNRs are not monitored in DNA-PAINT. In contrast, in the optoplasmonic sensor step-like signals are observed upon binding of the docking strands to GNRs. It has to be noted here that only a subset of all the docking strands attached to the GNRs provide step signals due to the different plasmonic enhancements at each binding site. 

\begin{figure*}[b!]
    \centering
    \includegraphics[width=\textwidth]{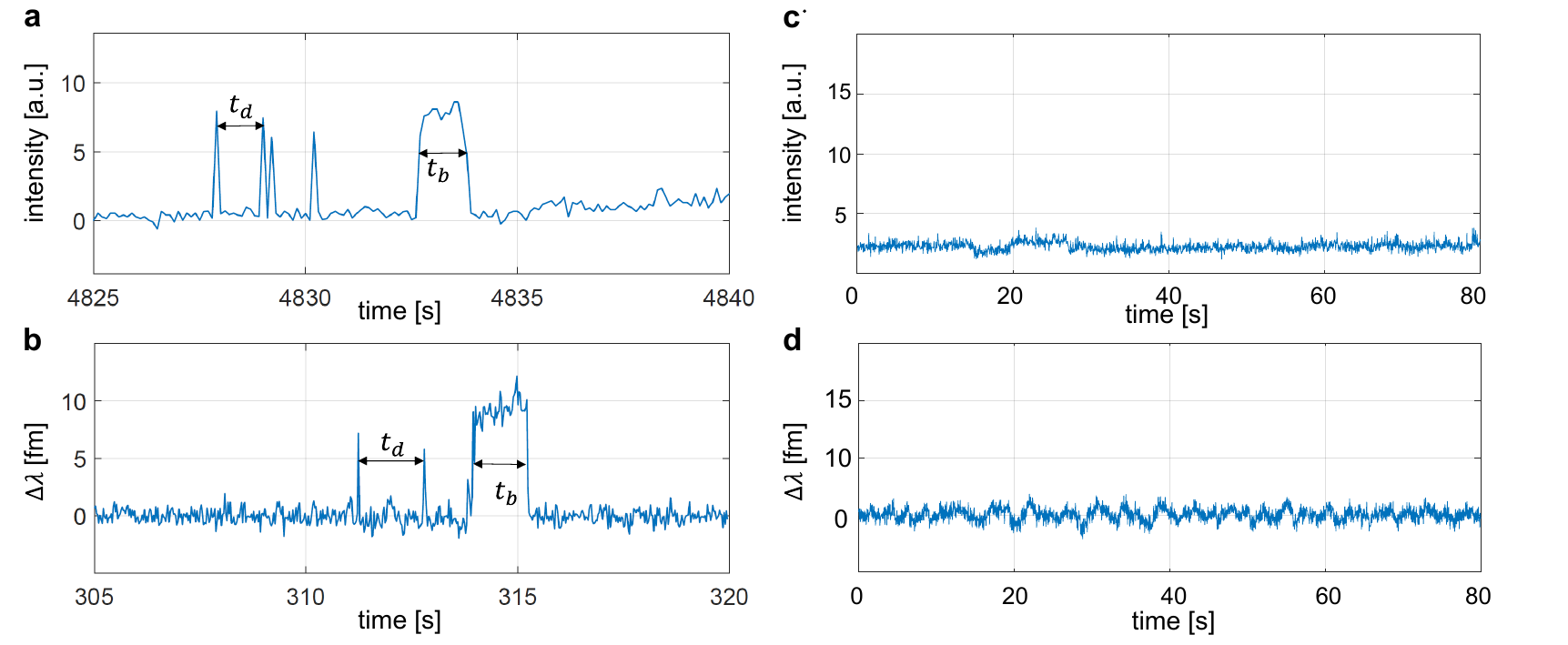} 
    \caption{(a,b) Examples of traces generated by the interactions between P1 docking and ImP1* imager strands in (a) the DNA-PAINT experiment, and (b) the optoplasmonic sensing technique. Similar traces can be observed for the T22 strands. Here $t_b$ denotes the time that a strand spends in a bounded state, while $t_d$ indicates the time between detected single-molecule hybridization events. (c,d) Control measurements of unrelated DNA strands for DNA-PAINT and WGM experiments, respectively. A lack of activity was observed in both cases, thus indicating that the observed DNA hybridization is specific to the sequences.}
    \label{fig:F2}
\end{figure*}

Finally, the transient interactions between the docking and imager strands are monitored in both techniques. After loading of the docking strands, the excess DNA in the sample chamber was removed by rinsing the chamber thrice with milli-Q water. The imager strands were then injected into the sample chamber at various concentrations. A buffer of Tris-EDTA (TE buffer, pH $\sim$ 8) with 500 mM NaCl was used for the experiments with P1 and its complementary, and a solution at pH $\sim$ 7 (milli-Q water) with 10 mM NaCl was used for the T22 DNA experiments. The different conditions were chosen to optimize the number of events observed to extract the various rates (P1 and T22 have very different melting temperatures\cite{breslauer1986predicting}). In DNA-PAINT, we then added the corresponding imager strands with dyes, i.e. ImP1*D modified with ATTO 655 (5' end) and ImT22*D with DY782 (5' end) for the docking strands P1 and T22, respectively. In each experiment, only one docking strand and the corresponding imager strand were used. The fluorescence signals from the hybridization of the DNA strands were recorded over time and co-localized with the position of the GNRs, as shown in Figure\,\ref{fig:F1}b (blue boxes). Figure\,\ref{fig:F1}c shows a space-time plot for the red line shown in Figure\,\ref{fig:F1}b. The constant intensity background indicates the GNR photoluminescence, and the burst of intensity indicates DNA hybridization between docking strands and imager strands that co-localized within GNR position. Fluorescent times traces can then be extracted using this information. An example is shown in Figure\,\ref{fig:F1}d. The blue curve shows the DNA hybridization between P1 and ImP1*D, and the orange curve displays the background signal adjacent to the GNR position. Along this time trace, most of the signals we observe are spike-like, although it is also possible to detect plateau-like signals (see Figure\,\ref{fig:F2}a) indicating prolonged interactions between imager and docking strands. In the case of the optoplasmonic sensor the imagers without fluorescent labels, ImP1* and ImT22* were used. Again, in each experiment only one set of docking and the corresponding imager strand was used and the imager concentrations was increased in steps. Upon addition of the imagers, the WGM resonance position shifts (see Figure\,\ref{fig:F1}g) and spike like transitions can be observed in a time trace of the WGM resonance position ($\Delta \lambda$). Figure\,\ref{fig:F1}h shows the spike signals due to interaction of the complementary DNA strands (blue) and the background with no spike signals before addition of the imager (orange).

DNA hybridization has previously been reported\cite{Jungmann2010,Andrews2021} to follow (pseudo) first-order kinetics. Considering a single docking strand that can be found in two states, either bounded with a single imager strand or dissociated, the probability that a binding event does not take place in an interval $t_d$ (dissociated waiting time), also called survival probability, is $P^s(t_d) = \exp{(-k_s t_d)}$, where $k_s$ is a single-molecule binding rate with units of $\rm{s^{-1}}$ (also known as association rate). However, the traces of both experiments (Figure\,\ref{fig:F2}) record the signal of binding events triggered by many imager strands, and the chance of observing the binding of one of the imagers in the chamber is further affected by the arrangement and accessibility of docking strands contributing to the signals on each sensor. The probability distribution for the waiting time to detect the binding of any one of the imagers, assuming independence of events, is then $P(t_d) = \exp{(-k t_d)}$, where $k = k_{\rm{on}} \gamma c_i$ is a rate with units of $\rm{s^{-1}}$. Here, $k_{\rm{on}}$ is the usual on-rate constant per unit of concentration, $\gamma$ is a numerical factor accounting for the geometric arrangement of the docking strands and their contribution to the observed signals, and $c_i$ is the imager concentration. Since the factor $\gamma$ is constant within the same experiment we expect that, for each experiment, the rate $k$ grows linearly with $c_i$. Once a docking strand is bound, the probability that dissociation does not take place in an interval $t_b$ (bound waiting time) is $P(t_b) \simeq \exp(-k_{\rm{off}} t_b),$ where $k_{\rm{off}}$ is the dissociation rate constant and, in this case, equivalent to a single molecule dissociation rate with units of $\rm{s^{-1}}$. In other words, we expect $k_{\rm{off}}$ to be concentration-independent.\cite{Jungmann2010,Andrews2021}  Note that the approximation holds provided that the chance of two imagers being bound at the same time is negligible, which is achieved by making the imager concentration sufficiently low. 

Using the values for the two waiting times, $t_d$ and $t_b$, all measured at temperature $22.5\,\rm{^oC}$, we computed the empirical distributions $P(t_d)$ and $P(t_b)$, both for the DNA-PAINT experiment (dotted line in Figures\,\ref{fig:F3}a,c) and the WGM platform (dotted line in Figures\,\ref{fig:F3}b,d). As expected, exponential profiles emerge, from which the rates $k$ and $k_{\mathrm{off}}$ are extracted for different molar imager concentrations $c_i$.

The single-molecule dissociation rate $k_{\rm{off}}$ can be directly obtained from the fits to $P(t_b)$ in Figure\,\ref{fig:F3}c (DNA-PAINT) and Figure\,\ref{fig:F3}d (WGM). As expected, the $k_{\rm{off}}$ values are indeed approximately constant across all trials with different imager concentrations $c_i$, as can be seen in Figure\,\ref{fig:F3}e-h (black lines). For P1 strands, the average dissociation rates are $k_{\rm{off}}^{\rm{PAINT}}= 3.6 \pm 0.4\,\rm{s^{-1}}$ and $k_{\rm{off}}^{\rm{WGM}} = 4.2 \pm 0.8\,\rm{s^{-1}}$ (estimate $\pm$ uncertainty; see caption of Fig.\,\ref{fig:F3}) for DNA-PAINT and WGM, respectively. Thus the value of $k_{\rm{off}}$ measured by both techniques is the same within error, suggesting that the fluorescent dye molecule in ImP1*D does not significantly alter the kinetics for the DNA hybridization under study. The $k_{\rm{off}}$ values for the T22 strands, $k_{\rm{off}}^{\rm{PAINT}}= 10 \pm 3\, \rm{s^{-1}}$ and $k_{\rm{off}}^{\rm{WGM}} = 9 \pm 1\, \rm{s^{-1}}$, lead to an analogous conclusion. Nevertheless, previous studies\cite{Andrews2021} have shown that the kinetics of biomolecular systems can be influenced by the presence of fluorophores, and so the impact of fluorophores may have to be examined in a case-by-case basis. The compatible $k_{\rm{off}}$ values also show that there is no evidence of a local temperature increase due to the near-field enhancement of the GNRs at the light intensities used in both techniques.

The relationship between $k$ and $c_i$ further provides the product $k_{\mathrm{on}}\gamma$. For P1 strands, a linear fit yields $k_{\rm{on}} \gamma^{\rm{PAINT}} = (8.7 \pm 3.4) \times 10^7\,\rm{M^{-1} s^{-1}}$  and $k_{\rm{on}} \gamma^{\rm{WGM}} = (3.9 \pm 0.5) \times 10^4\,\rm{M^{-1} s^{-1}}$ for the DNA-PAINT and optoplasmonic experiments, respectively. Similarly, for T22 strands we find $k_{\rm{on}} \gamma^{\rm{PAINT}} = (1.3 \pm 0.4)\times 10^7\,\rm{M^{-1} s^{-1}}$ and $k_{\rm{on}} \gamma^{\rm{WGM}} = (1.1 \pm 0.1)\times 10^6\,\rm{M^{-1} s^{-1}}$. These fits are shown in Figures\,\ref{fig:F3}e-h (orange lines), and, for both techniques, we see clear evidence of a linear dependence between $k$ and $c_i$. The value of $k_{\rm{on}}$ for each strand cannot be determined without detailed knowledge of the geometric factors. An exact calculation of $\gamma$ for each technique lies beyond of the scope of the present work, but we can highlight some factors contributing to its value. First, all docking strands contribute to the signal in DNA-PAINT, whereas only $12\%$ of the docking strands in the WGM technique (attached to the tips of the GNRs) contribute to the signal. This already suggests that $k_{\rm{on}}\gamma$ can be smaller for the WGM sensor than for DNA-PAINT, in consistency with our measurements. Additionally, the local surface properties will be different for the tips compared to the sides of the GNRs \cite{Willets2019,Byers2014,Kaushal2020} Of particular importance is the fact that, in the WGM experiment, the plasmonic enhancement may be insufficient for many docking sites due to the random deposition of the GNRs on the surface of the microsphere, thus failing to detect signals that lie below the noise level. For P1 strands, this is reflected in that only when applying a micromolar concentration of imager strands can one detect a significant number of events. In contrast, for T22 strands the number of events is significant already in the range of hundreds of $\rm{nM}$. This gives a plausible explanation of why the discrepancy between $k_{\rm{on}} \gamma^{\rm{PAINT}}$ and $k_{\rm{on}} \gamma^{\rm{WGM}}$ is only of $\sim 10$ for the T22 strands, but of $\sim 10^3 - 10^4$ for the P1 strands. 

Here we have compared two optical single-molecule detection techniques, fluorescence-based single-molecule localization (DNA-PAINT) and optoplasmonic WGM sensing, that utilize the same plasmonic nanoparticles for detecting DNA hybridization events between surface immobilized docking strands and imager strands in solution. The rate constant $k_{\rm{off}}$, which characterizes the dissociation of the DNA sequences used here, should depend only on the oligonucleotide sequence and length (buffer and temperature conditions are the same for each type of DNA across both experimental setups) while being independent of the different geometric factors of the two experiments. Indeed, we found $k_{\rm{off}}$ to be the same in both experiments, within experimental error. This establishes the equivalence of each technique for test systems used in this work and can serve as the basis for consistently combining these techniques in future single-molecule studies. In particular, to establish the dissociation rates for other processes, one may benefit from the larger binding efficiency of DNA-PAINT, while checking the label-impact with the WGM platform. 	 

\section*{Author Information}

\subsection*{Corresponding Author}
\url{*ne276@exeter.ac.uk }

\subsection*{Author Contributions}
N.E. and J.R. analyzed the data and wrote the manuscript.; S.S. carried out the WGM-based experiment; N.E., T.L. and S.S. conducted the DNA-PAINT experiments; N.E.,T.L. and C.S. analyzed the raw DNA-PAINT data; H.-Y.W. trained N.E. in conducting experiments; J.A. advised on the data analysis and supervised the project; C. S. and F.V. conceived the study. All authors contributed to writing the manuscript.  
\subsection*{Notes}
The authors declare no competing financial interests.

\section*{ACKNOWLEDGMENT}
N.E. acknowledges funding from the EPSRC Centre for Doctoral Training (CDT) in Metamaterials (XM2) at the University of Exeter. N.E., H.-Y.W. and F.V. acknowledge support from EPSRC (EP/R031428/1). S.S., J.R., J.A. and F.V. acknowledge support from EPSRC (EP/T002875/1). J.A. acknowledges support from EPSRC (EP/R045577/1) and the Royal Society. C.S. acknowledges support from EPSRC (EP/N008235/1).

\bibliographystyle{unsrt}  
\bibliography{references}  
\end{document}